\documentclass[a4paper,11pt]{article}
\usepackage{pos}

\title{Muon g-2}
\author*[a]{Christine Davies}
\affiliation[a]{School of Physics and Astronomy,\\
 University of Glasgow, Glasgow, G12 8QQ, U.K.}

\emailAdd{Christine.davies@glasgow.ac.uk}
\abstract{The story of the anomalous magnetic moment of the muon, $a_{\mu}$, has been a long and absorbing read, with several unexpected plot twists. The theme: is new physics visible in $a_{\mu}$ or not? is clearly an important one. As we head towards what {\it may} be the final chapters, the pace of the narrative has quickened and lattice QCD is providing a critical new perspective on some of the key characters. I will review the story so far, focussing on recent episodes but ({\it spoiler alert}) depending on your point of view, it may not be heading to a happy ending. }

\FullConference{The 41st International Symposium on Lattice Field Theory (LATTICE2024)\\
 28 July - 3 August 2024\\
Liverpool, UK\\}

\begin{document}
\maketitle

\section{Introduction}
\label{sec:intro}
The muon has electric charge and spin and hence a magnetic moment. This is related to its spin by 
\begin{equation}
\label{eq:magdef}
\vec{\mu}_{\mu}=g_{\mu}\left(\frac{e}{2m_{\mu}}\right) \vec{S}\, .
\end{equation}
The $g$-factor would be 2 in a purely quantum-mechanical world. However, quantum field theory provides a rich vacuum structure with a sea of virtual particles. As the muon spins, its magnetic moment probes the vacuum and this increases the value of $g$ by about 0.1\%. The anomalous magnetic moment, or simply `anomaly', is defined as 
\begin{equation}
\label{eq:adef}
a_{\mu}=\frac{g_{\mu}-2}{2} \, .
\end{equation}

\begin{figure}[thb]
  \centering
  \includegraphics[width=0.25\hsize]{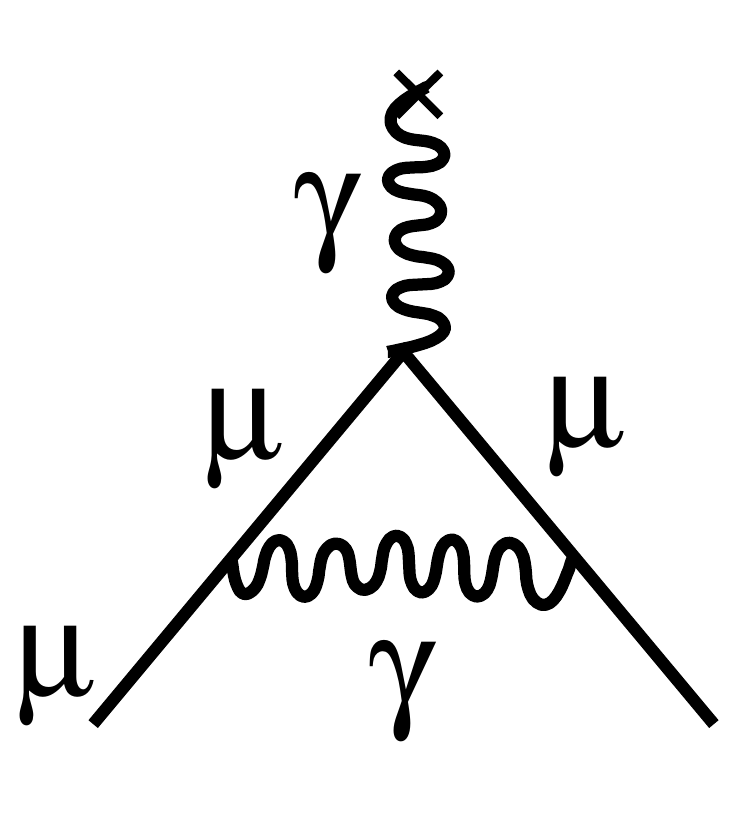}
  \hspace{5.0em}
    \includegraphics[width=0.28\hsize]{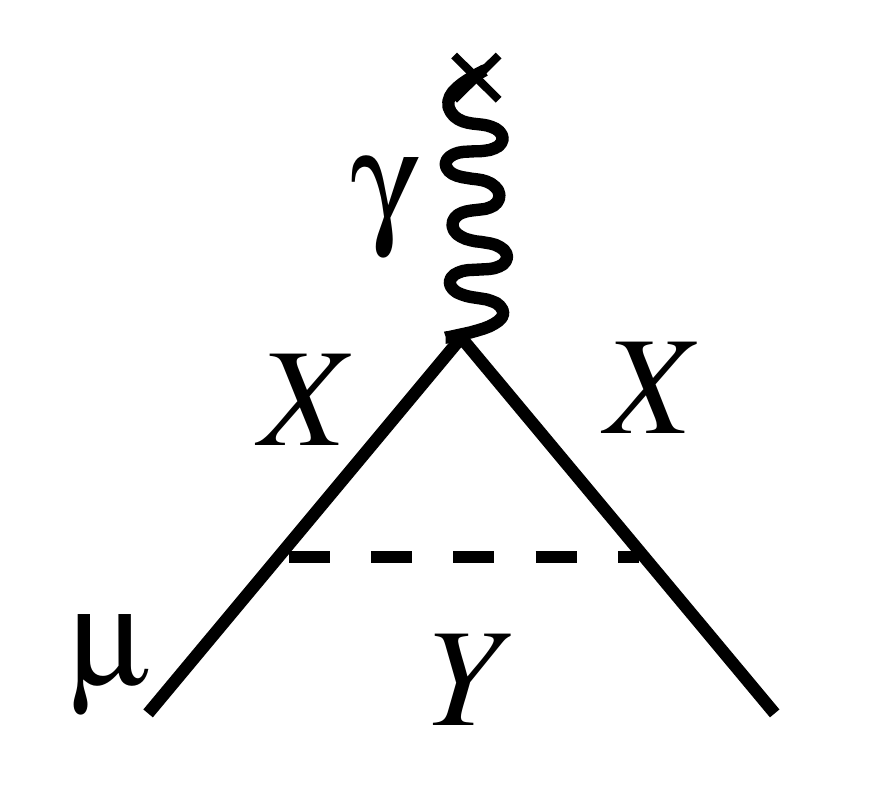}
  \caption{The left-hand plot shows the leading-order contribution to $a_{\mu}$, from QED, equal to $\alpha/(2\pi)=0.00116$. The vertical photon represents the $B$ field and the top $\mu-\gamma$ vertex then defines the $\mu$ magnetic moment. The right-hand plot shows a possible new physics contribution.}
  \label{fig:LO-NP}% Give a unique label
\end{figure}

The leading contribution to $a_{\mu}$ comes at $\mathcal{O}(\alpha)$ from QED (see Figure~\ref{fig:LO-NP}), but there are many higher-order pieces. The task of the theorist is to calculate them all. A discrepancy between $a_{\mu}$ calculated in the Standard Model (SM) and that determined in experiment would constitute a discovery of new physics. A possible new physics diagram is shown on the right of Figure~\ref{fig:LO-NP}. 

$a_{\mu}$ is being determined experimentally by the Muon g-2 collaboration at Fermilab to an astonishing level of precision. In August 2023 the collaboration released an update~\cite{Muong-2:2023cdq} based on data runs 1--3 which gave an experimental world average with an accuracy of 0.19ppm. The experiment is in fact done anti-muons, spin-polarised and circulating in a storage ring with a perpendicular magnetic field. Since the ${\mu}^+$ decay weakly (to $e^+\overline{\nu}_{\mu}\nu_e$) their spin direction can be measured from that in which the high-energy positrons are emitted. The precession of the ${\mu}^+$ spin is then compared to that of its linear momentum. The precession frequency difference is directly proportional to $a_{\mu}$, avoiding the need to measure the `2' in the numerator of Eq.~\eqref{eq:adef} and making high precision feasible. A final result is expected in 2025 from runs 1--6, improving statistical uncertainties by a further factor of 3 and the total uncertainty by 30\%. 

The current experimental value can be compared to that expected in the Standard Model, as given in the 2020 Theory White Paper (WP20)~\cite{Aoyama:2020ynm}: 
\begin{eqnarray}
\label{eq:values}
10^{10}a_{\mu}^{\mathrm{expt}} &=& 11659205.9(2.2)  \nonumber \\
10^{10}a_{\mu}^{\mathrm{theo}} &=& 11659181.0(4.3)  \, .
\end{eqnarray}
The difference between these two numbers, $24.9(4.8)$ would constitute a $5\sigma$ discovery, except that it has become clear since 2020 that the QCD contributions need more work. The relative size of contributions from different sectors of the SM, as taken from WP20 (and adding up to the $a_{\mu}^{\mathrm{theo}}$ given in Eq.~\eqref{eq:values}) are: 
\begin{eqnarray}
\label{eq:contribs}
\mathrm{QED} (\times10^{10}) &=& 11658471.8931(104)  \nonumber \\
\mathrm{EW}  (\times10^{10}) &=& 15.36(10) \nonumber \\
\mathrm{QCD} (\times10^{10}) &=& 693.7(4.3) \,.
\end{eqnarray}
These numbers show that the QCD contribution is the second largest, much larger than that from EW effects. This is because the corrections to $g$ come from low energies, around $m_{\mu}$, and the EW gauge bosons are very heavy. The QED and EW contributions can be calculated accurately in perturbation theory. The QCD contribution can not and so it gives almost the entire theory uncertainty. Lattice QCD then has a key role to play in clarifying what that contribution is. 

\section{QCD contributions to $a_{\mu}$}
\label{sec:QCD}

\begin{figure}[thb]
  \centering
  \includegraphics[width=0.8\hsize]{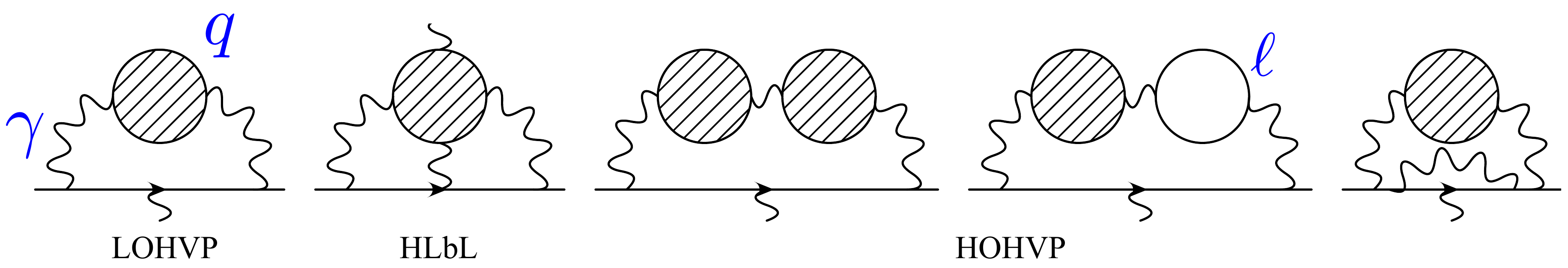}
  \caption{QCD contributions to $a_{\mu}$. From left to right: the leading-order hadronic vacuum polarisation (LOHVP), hadronic light-by-light (HLbL), higher-order HVP (HOHVP). The horizontal lines represent the $\mu$ and wiggly lines are photons, with the vertical photon the $B$ field. Filled loops are strongly-interacting and formed from $\gamma^*\to q\overline{q}\to\gamma^*$; open loops are lepton loops. Figure from~\cite{Blum:2012jdu}.}
  \label{fig:QCD}% Give a unique label
\end{figure}

The leading (in powers of $\alpha$) QCD contribution is that on the left of Figure~\ref{fig:QCD}, known as the LOHVP, which appears at $\mathcal{O}(\alpha^2)$.  The remaining four diagrams are $\mathcal{O}(\alpha^3)$ and consequently give a much smaller contribution. The total for each of the HOHVP and HLbL collections of diagrams is about a factor of 70 smaller than the LOHVP~\cite{Aoyama:2020ynm}. There are additional $\alpha^4$ versions of these diagrams that are even smaller (but are included in the total in~\cite{Aoyama:2020ynm}). The HOHVP contribution is negative and almost cancels that of the HLbL. The total QCD contribution is then dominated by that of the LOHVP. For a theory uncertainty of $\sim 3.5\times 10^{-10}$ (i.e. in the same ballpark as the experiment) we need a 0.5\% uncertainty on the LOHVP calculation. The HOHVP result inherits a similar relative uncertainty whilst being much smaller, so its uncertainty is not problematic. The HLbL diagram is hard to pin down and so historically it has had a large relative uncertainty. If this can be reduced to $\mathcal{O}(10\%)$ then its uncertainty will be small compared to that of the LOHVP. 

\begin{figure}[thb]
  \centering
  \includegraphics[width=0.8\hsize]{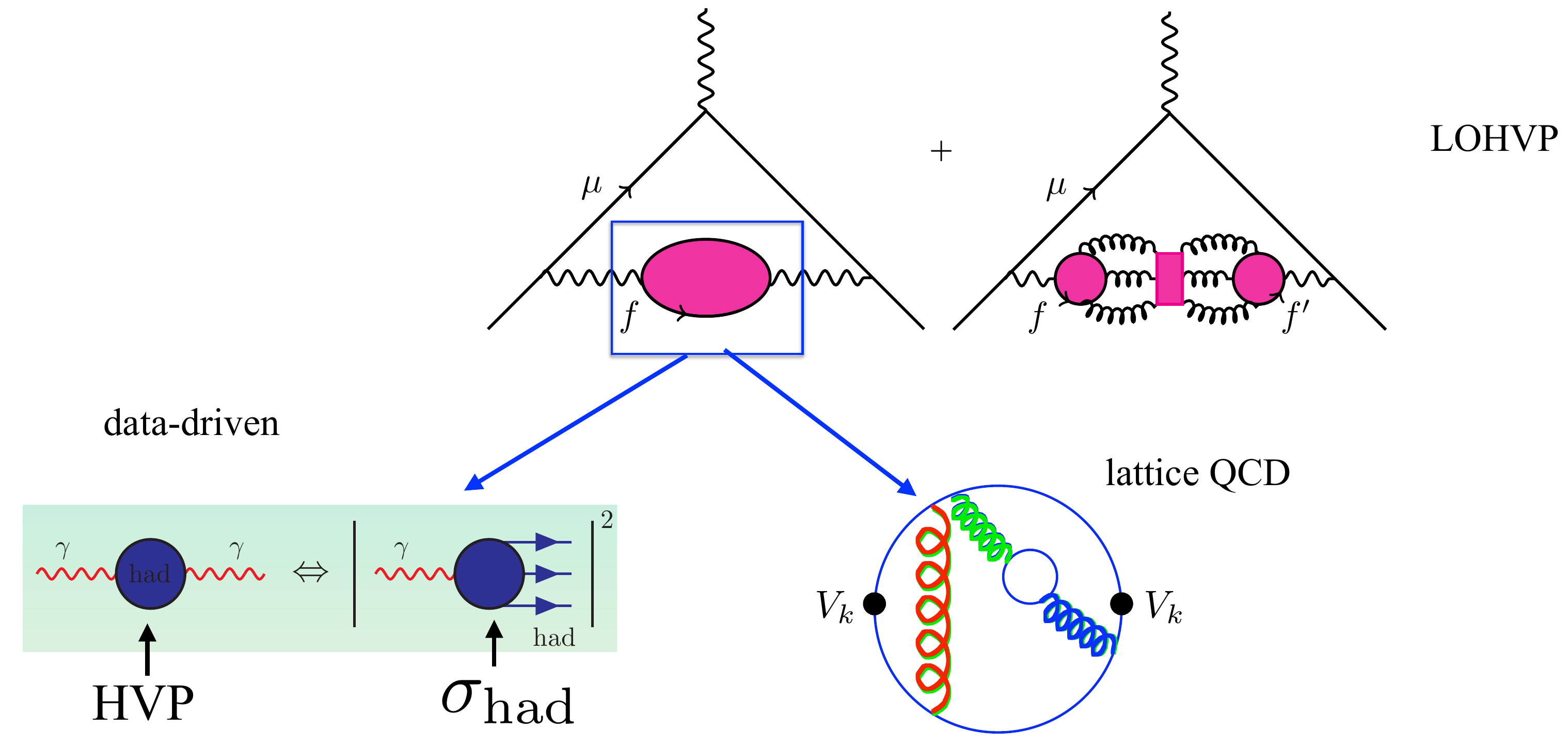}
  \caption{Illustrating the two methods for determining the LOHVP contribution to $a_{\mu}$. The basic LOHVP diagram is that on the top left with a quark bubble of flavour $f$ attached to a photon on either side. As part of this, however, we must also consider the quark-line disconnected diagram on the right (where $f$ and $f^{\prime}$ can be different flavours) as well as QED corrections internal to the quark bubble. Concentrating for illustrative purposes on the simple quark bubble, the two different methods are shown below. The lattice QCD approach calculates the quark bubble as a simple correlation function on the lattice ($V_k$ is the electromagnetic current). In the data-driven approach the bubble is converted, via analyticity and the optical theorem, into an expression involving the cross-section for a virtual photon (produced in $e^+e^-$ collisions) to decay to hadrons. The data-driven approach automatically includes quark-line disconnected and QED effects; in lattice QCD they must be calculated separately. }
  \label{fig:hvpmethods}% Give a unique label
\end{figure}

Here we will largely concentrate on the calculation of the LOHVP contribution; that contribution is clearly critical to providing an accurate theoretical result for $a_{\mu}$ in the SM. There are two approaches to calculating the LOHVP, illustrated in Figure~\ref{fig:hvpmethods}. The `data-driven' approach that makes use of experimental data for $\sigma(e^+e^-\to \mathrm{hadrons})$ has historically been the most accurate and this gave the value quoted in WP20~\cite{Aoyama:2020ynm} and Eq.~\eqref{eq:contribs}. The lattice QCD calculation required is that of simple 2-point vector-vector correlation functions for different quark flavours but it has proved very hard to achieve the accuracy needed. 

\begin{figure}[thb]
  \centering
  \includegraphics[width=0.6\hsize]{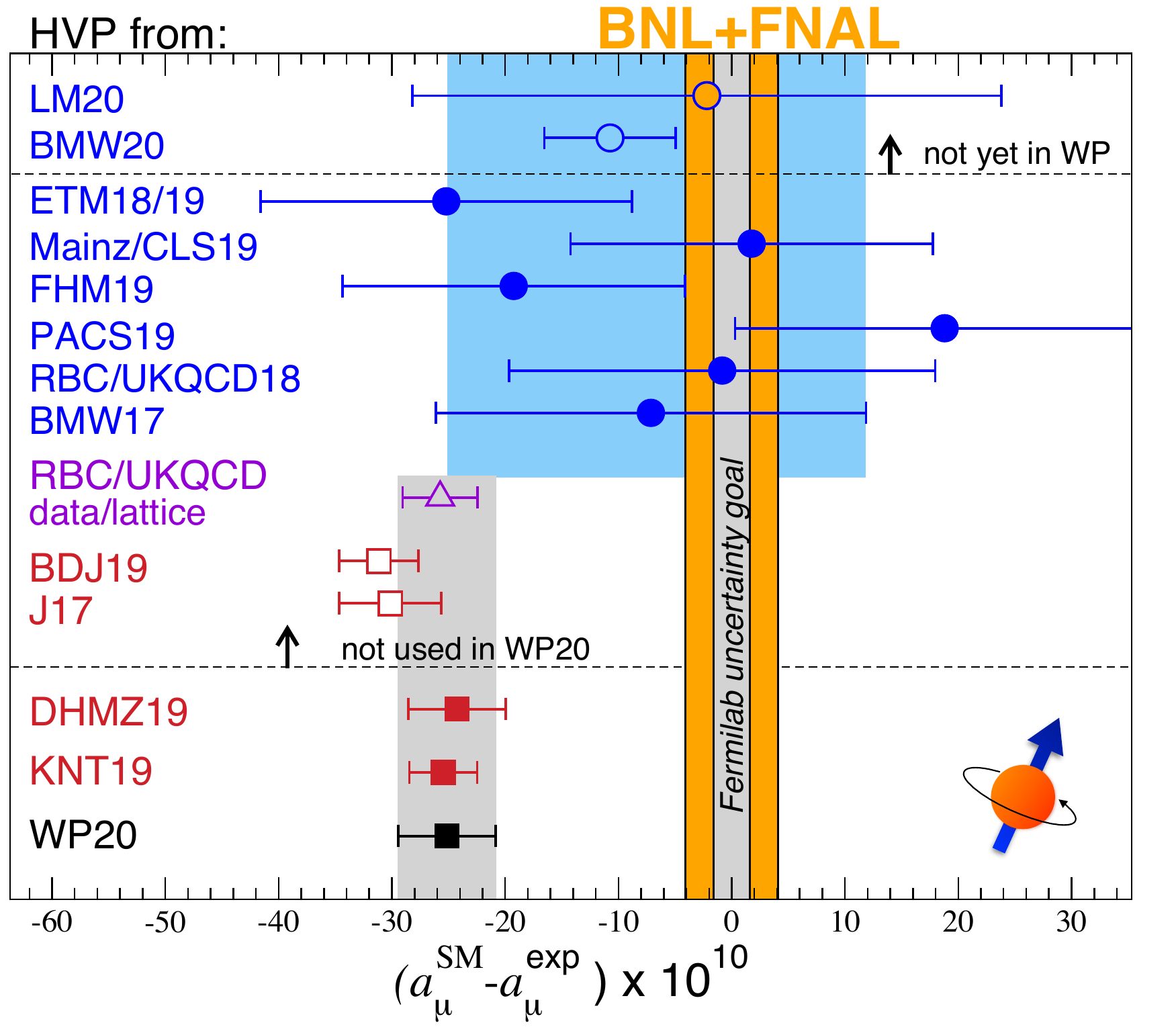}
  \caption{Comparison of theoretical predictions for $a_{\mu}$ with experiment from the 2022 Snowmass review~\cite{Colangelo:2022jxc}. Experiment is shown as the orange band; the uncertainty has since been halved~\cite{Muong-2:2023cdq}. Each data point represents a different evaluation of the LOHVP to which the same value for the other contributions has been added from WP20~\cite{Aoyama:2020ynm}. Red squares indicate the data-driven results~\cite{davier:2019can,keshavarzi:2019abf} that were combined to give the WP20 value given by the black square and grey band. This grey band shows a tension of $\sim 25(5)$ with the orange band, as discussed following Eq.~\eqref{eq:values}. The lattice QCD results available at that time are shown in blue, with the filled blue circles~\cite{Budapest-Marseille-Wuppertal:2017okr,RBC:2018dos,Giusti:2019xct,Shintani:2019wai,FermilabLattice:2019ugu,Gerardin:2019rua,Giusti:2019hkz} giving the WP20 lattice average as the blue band. The two open blue circles are lattice results~\cite{Borsanyi:2020mff,Lehner:2020crt} published after the WP20 deadline. }
  \label{fig:snowmass}% Give a unique label
\end{figure}

The situation in early 2022 was summarised by the Snowmass review~\cite{Colangelo:2022jxc}, Figure~\ref{fig:snowmass}. This shows the impact of the value of the LOHVP contribution on the comparison of $a_{\mu}$ between theory and experiment. The significant tension of Eq.~\eqref{eq:values} between experiment (orange band) and theory (grey band), using the WP20 LOHVP value based on data-driven results, is evident. The WP20 LOHVP value had a 0.6\% uncertainty. Note that the size of the tension, $25(5)\times 10^{-10}$, is $\sim3.5\%$ of the LOHVP. The lattice QCD values available at that time mostly had uncertainties at the 2\% level and the WP20 lattice average, given by the blue band, was unable to provide the resolution needed to weigh in on the issue. The BMW20 result~\cite{Borsanyi:2020mff}, not included in the WP20 lattice average, was the first complete lattice QCD calculation of the LOHVP and has a 0.8\% uncertainty. The $2\sigma$ tension that it shows with the data-driven LOHVP values, and less than $2\sigma$ tension with experiment, was intriguing. It demonstrates that the question of whether the data-driven and lattice QCD values for the LOHVP agree is critical to forming a conclusion on the evidence for new physics in $a_{\mu}$. 

\section{Windows on the lattice LOHVP calculation}
\label{sec:latwindows}
\begin{figure}[t]
  \centering
  \includegraphics[width=0.8\hsize]{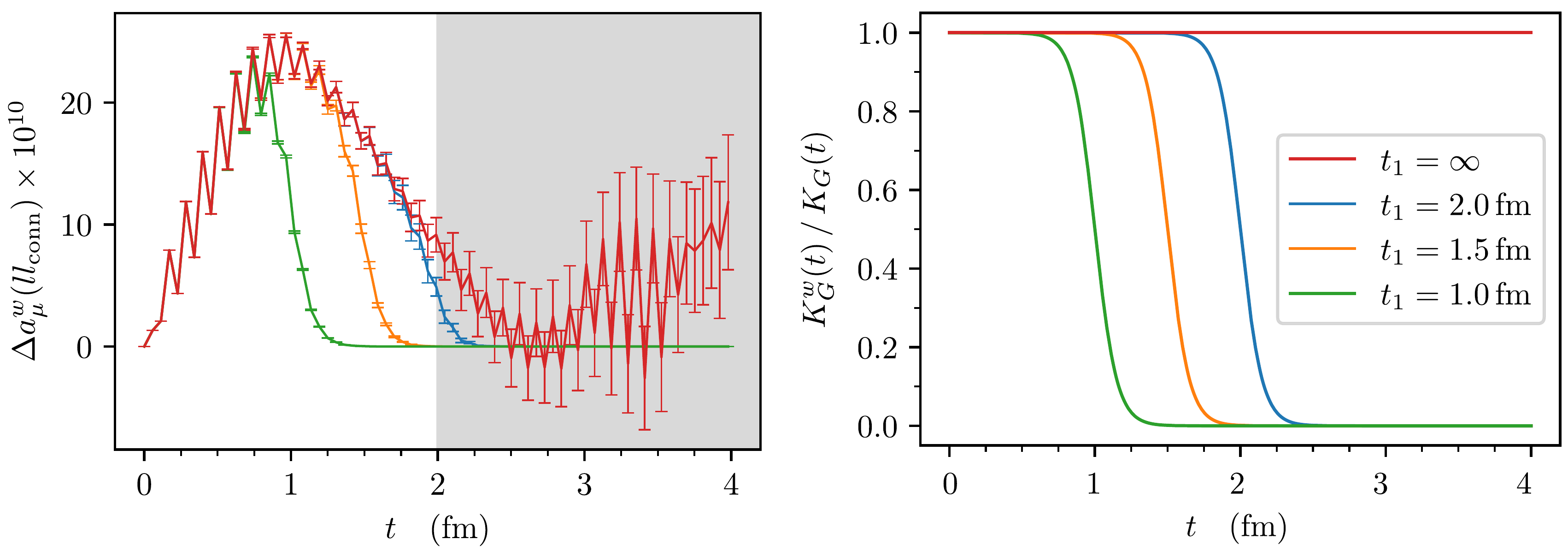}
  \caption{The left-hand plot shows in red the contributions to $a_{\mu}^{\mathrm{LOHVP}}$ as a function of time, i.e. the product $G(t)\tilde{K}(t)\alpha^2/\pi^2$ of Eq.~\eqref{eq:lattLOHVP}, for the connected light-quark correlation function calculated using HISQ quarks (the correlator oscillations are an inconsequential effect from staggered quarks). Note how the statistical uncertainties grow with $t$; the results were deemed to be unreliable in the grey shaded region above $t=2\,\mathrm{fm}$. The right-hand plot shows the one-sided window functions for different $t_1$ (Eq.~\eqref{eq:window}) that are multiplied into the integrand of Eq.~\eqref{eq:lattLOHVP} to produce partial results from the lattice calculation with smaller statistical errors. The left-hand plot shows the impact of these one-sided windows on the integrand, with the same colour coding as on the right. Figures from~\cite{FermilabLattice:2022izv}.  }
  \label{fig:onesidedwindows}% Give a unique label
\end{figure}

As shown in Fig.~\ref{fig:hvpmethods} the lattice QCD calculation of the LOHVP requires a straightforward determination of 2-point vector-vector correlation functions at zero spatial momentum, i.e. the calculation that would be done to determine vector meson masses. We must also determine quark-line disconnected correlation functions along with QED and isospin-breaking effects. As usual results are needed at multiple values of the lattice spacing for a continuum limit to be taken. Given an appropriate correlation function, $G(t)= \langle V_k(0)V^{\dag}_k(t)\rangle$, the contribution to the LOHVP is given by 
\begin{equation}
\label{eq:lattLOHVP}
a_{\mu}^{\mathrm{LOHVP}}=\left(\frac{\alpha}{\pi}\right)^2 \int_0^{\infty} dt G(t) \tilde{K}(t) \, . 
\end{equation}
Here $V_k =\sum_f e_f \overline{\psi}_f \gamma_k \psi_f$ and  $\tilde{K}$ is a kernel function~\cite{Blum:2002ii,Bernecker:2011gh} which vanishes at $t=0$, and rises up from that value as $G(t)$ falls. The product $G(t)\tilde{K}(t)$ for quark-connected correlators then has a humped structure with the position of the peak of the hump depending on the quark flavour. This is illustrated in Figure~\ref{fig:onesidedwindows} for the connected light quark ($m_u=m_d=m_l$) correlator in red calculated with Highly Improved Staggered quarks~\cite{FermilabLattice:2022izv}. Heavier quark contributions ($s,c$) peak at smaller $t$. The size of the contribution from each quark flavour then depends on the effective $t$ range over which we need to sum for Eq.~\eqref{eq:lattLOHVP} as well as the quark electric charge, which appears squared in $G(t)$. 

\begin{figure}[thb]
  \centering
  \includegraphics[width=0.8\hsize]{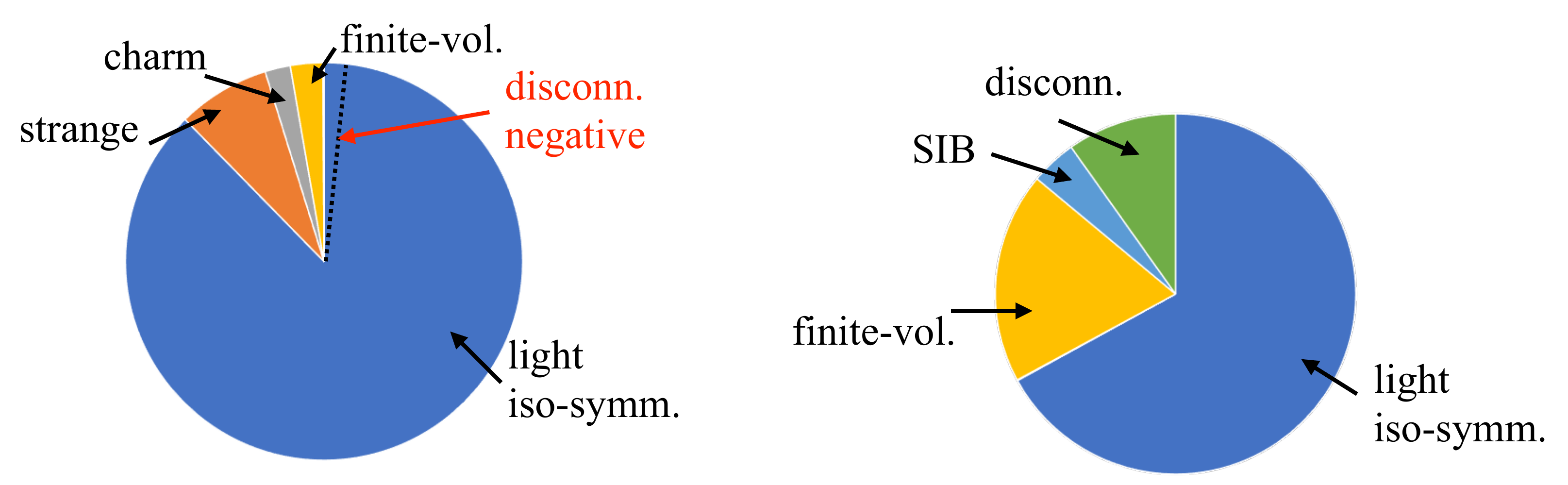}
  \caption{Pie charts showing the relative size of different contributions to the LOHVP (left) and to its variance (right) from lattice QCD calculations in~\cite{Borsanyi:2020mff}. Dark blue shows the light iso-symmetric ($m_u=m_d=m_l$) contribution which dominates both.  The size of the negative quark-line disconnected contribution is shown, via the dotted line, as part of the light iso-symmetric contribution. }
  \label{fig:lattice-full-pie}% Give a unique label
\end{figure}

The relative sizes of the different flavour contributions are shown in the left-hand pie chart of Figure~\ref{fig:lattice-full-pie}, with numbers taken from BMW20~\cite{Borsanyi:2020mff}. We see that by far the largest contribution ($\sim 90\%$) comes from up and down quarks, with strange (partly because of $e_s$) and charm (because of its mass) having much smaller contributions. The quark-line disconnected contribution (from $u,d,s$) is negative and has about the same magnitude as the charm contribution. The finite-volume correction is also sizeable (for a lattice of spatial size 6.272 fm used as the reference in this case). Contributions from QED and isospin-breaking as well as $b$ quarks are too small to be visible (but still need to be calculated, of course).  The right-hand pie shows the contributions to the variance, again dominated by the light-quark connected contribution. The uncertainty from the finite-volume corrections is significant, however. 

\begin{figure}[thb]
  \centering
  \includegraphics[width=0.9\hsize]{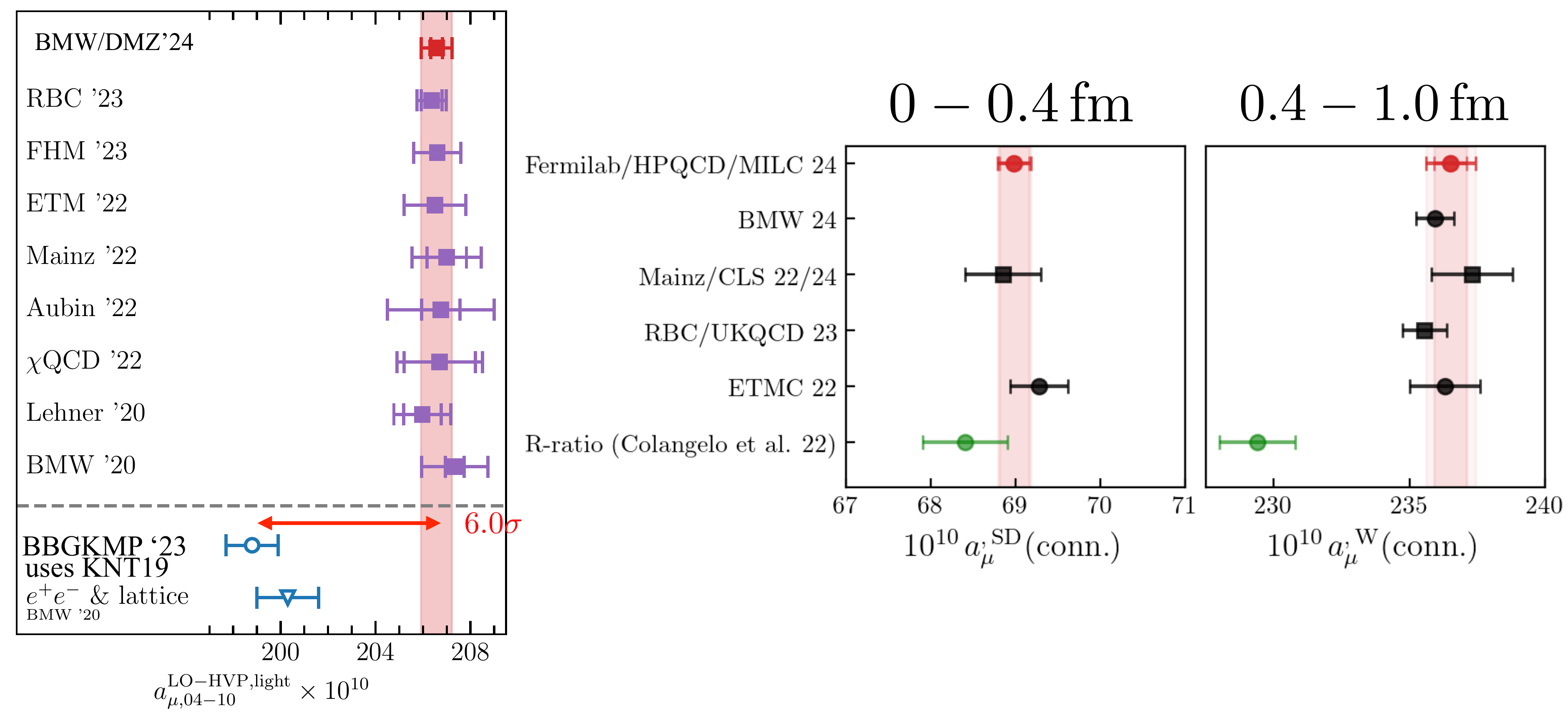}
  \caption{Left: a comparison of lattice QCD results for the light-quark connected contribution to the LOHVP in the intermediate distance window (0.4--1.0 fm). The figure is from BMW/DMZ24~\cite{Boccaletti:2024guq} and the red square and pink band gives their updated result, which has a 0.3\% uncertainty. Earlier lattice QCD results~\cite{Borsanyi:2020mff,Lehner:2020crt,Wang:2022lkq,Aubin:2022hgm,Ce:2022kxy,ExtendedTwistedMass:2022jpw,RBC:2023pvn,FermilabLatticeHPQCD:2023jof} are shown as purple squares and use six different quark formalisms between them. For comparison, the open blue triangle~\cite{Borsanyi:2020mff} uses the lattice calculation to remove other quark flavour contributions from the data-driven value corresponding to WP20. The open blue circle~\cite{Benton:2023fcv} uses an isospin analysis of different channels in the $e^+e^-$ data~\cite{keshavarzi:2019abf} to isolate the $I=1$ piece that corresponds to the light-quark connected contribution. There is a $6\sigma$ (3.7\%) difference between BMW/DMZ24 and this data-driven result based on pre-CMD3 experimental data. \\ Right: a comparison of lattice QCD results for the complete contribution to the LOHVP in the short-distance (0--0.4 fm) and intermediate distance windows. The figure is from Fermilab/HPQCD/MILC (FHM) and the top filled red circles and pink bands are their results from~\cite{FermilabLattice:2024yho} with uncertainties of 0.3\% ad 0.4\% respectively. The other lattice results~\cite{Ce:2022kxy,ExtendedTwistedMass:2022jpw,RBC:2023pvn,Boccaletti:2024guq} are in good agreement. Results from the pre-CMD3 data-driven analysis~\cite{Colangelo:2022vok} are shown as filled green circles. There is significant tension ($4\sigma$) with lattice QCD for the intermediate-distance window. When the CMD3 data for $e^+e^-\to \pi^+\pi^-$ is substituted for the older data in the KNT19 dataset the partial LOHVP values increase; for the short-distance window to 69.1 and for the intermediate-distance window to 235.6 (using numbers from~\cite{Davies:2024pvv}). These numbers are in better agreement with the lattice QCD results. This substitution of CMD-3 data also removes the tension between lattice QCD and data-driven results in the light-quark connected case~\cite{Benton:2023fcv}. }
  \label{fig:ID-comp}% Give a unique label
\end{figure}

An accurate LOHVP calculation then requires good control of uncertainties from the light-quark connected correlators. As is well-known, vector meson correlators have an exponentially falling signal/noise ratio at large times because of the gap between the vector meson mass (in the signal) and the pseudoscalar mass (controlling the noise). This affects the light meson correlation function more than that for $s$ or $c$ because the gap, $\pi$ to $\rho$, is larger. The integrand of Eq.~\eqref{eq:lattLOHVP} also extends to larger times for the light correlator because the exponential decay of G(t) is not so rapid and this exacerbates the statistical noise issue. The growth of statistical noise at large time is clearly visible in the red points in the left-hand plot of Figure~\ref{fig:onesidedwindows}. Another important issue for the light-quark case, especially at physical light quark masses, is that of finite-volume corrections. These arise from the fact that the light vector correlator has $\pi-\pi$ contributions, which become increasingly important at large time. S. Kuberski reviewed these sources of uncertainty and mitigation strategies for them at Lattice2023~\cite{Kuberski:2023qgx}. 

As we saw in Figure~\ref{fig:snowmass}, lattice QCD calculations were struggling to have an impact because of these uncertainties. It was suggested~\cite{Bernecker:2011gh,RBC:2018dos} that imposing a time-window on the lattice correlators to remove the problematic large time region where statistical and systematic uncertainties are largest would enable a more focussed comparison of lattice results for part of the LOHVP. Figure~\ref{fig:onesidedwindows} shows examples of such windows given by the function 
\begin{equation}
\label{eq:window}
0-t_1\,\,\, \mathrm{window}: \quad \theta(t,t_1,\Delta t) = \frac{1}{2}\left[ 1 - \tanh\left(\frac{t-t_1}{\Delta t}\right) \right] \, .
\end{equation}
This cuts out time values above $t_1$ with a rounded edge set by $\Delta t$, taken to be 0.15 fm. The left-hand plot shows the impact this has on the integrand and it can easily be seen that the partial LOHVP results will be much more precise. The standard windows that have been generally adopted use adaptions of this function to divide the time region into three: the short-distance window from 0--0.4 fm, the intermediate-distance window from 0.4--1.0 fm and the long-distance window above 1.0 fm (see Figure~\ref{fig:Rswindows}, which also shows in the upper pie charts the flavour breakdown of each window). The windowed results can simply be added together to obtain the total. 

Figure~\ref{fig:ID-comp} shows lattice results for the intermediate-distance window from light-quark connected correlators only (on the left) and the complete contribution for the short-distance and intermediate-distance windows (on the right). We notice immediately how much more impressive they are than the lattice results in Figure~\ref{fig:snowmass}. There is striking agreement on results for the intermediate window from multiple different lattice groups using different quark formalisms with uncertainties at the level of 1\% or below. Note that it is standard practice now to `blind' results until the analysis is finalised to remove any bias from `knowing' the answer. In fact these results constitute one of the best tests of full lattice QCD that has been done.

\section{Windows on the data-driven LOHVP calculation}
\label{sec:datawindows}
\begin{figure}[t]
  \centering
  \includegraphics[width=0.9\hsize]{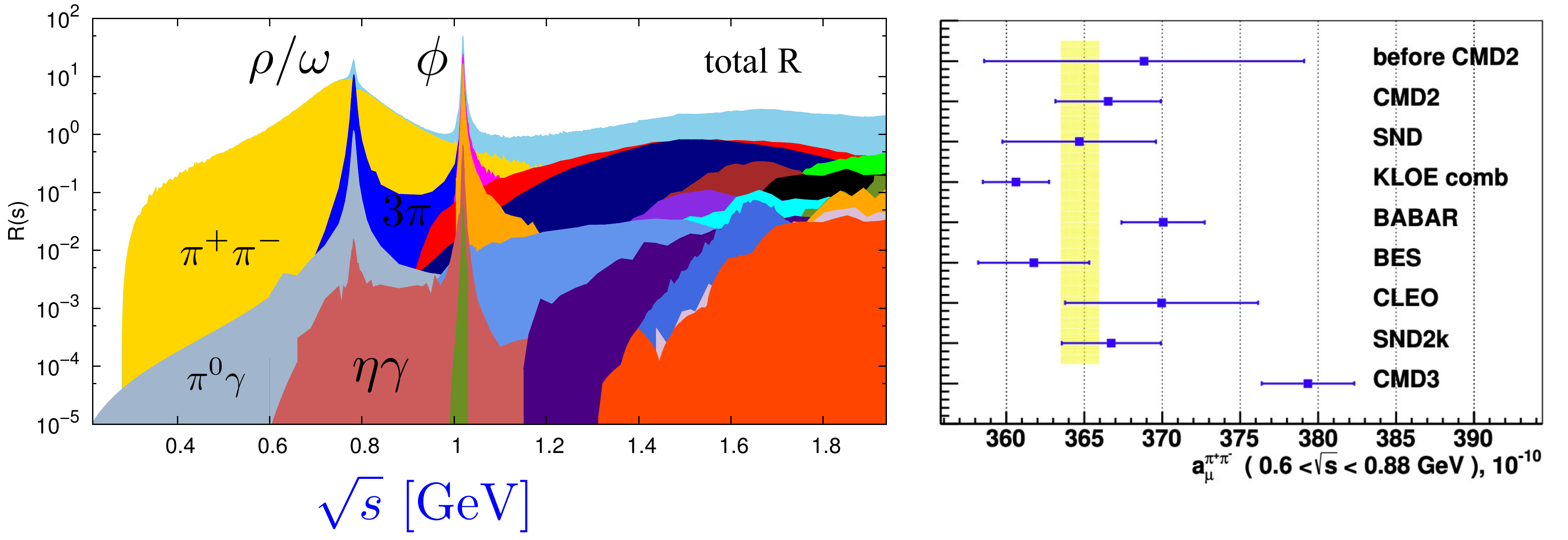}
  \caption{The left-hand plot shows the experimental results for different channels contributing to the total value for $R(s)$ at low values of centre-of-mass-energy, $\sqrt{s}$. $R(s)$ is the ratio of $\sigma(e^+e^-\to\mathrm{hadrons})$ to $4\pi\alpha^2/(3s)$. Note the logarithmic y-axis scale. Figure from A. Keshavarzi. The right-hand plot shows the tension between results for different experiments for the contribution to the LOHVP from $e^+e^-\to\pi^+\pi^-$ cross-section data. The recent CMD-3 results~\cite{CMD-3:2023alj,CMD-3:2023rfe} (lowest point) double the tension that previously existed. The yellow band shows the average of the experimental results before CMD-3. Figure from~\cite{CMD-3:2023alj}. Other experimental data shown are from~\cite{akhmetshin:2003zn,aulchenko:2006na,akhmetshin:2006wh,akhmetshin:2006bx,achasov:2006vp,ambrosino:2008aa,ambrosino:2010bv,babusci:2012rp,anastasi:2017eio,lees:2012cj,ablikim:2015orh}.  }
  \label{fig:datadriven}% Give a unique label
\end{figure}

We now have accurate lattice QCD results for the partial LOHVP and we turn to a brief discussion of the data-driven approach with which we need to compare them. This  was reviewed at Lattice 2023 by A. Keshavarzi. The method takes experimental data for $\sigma(e^+e^- \to \mathrm{hadrons})$ as a function of centre-of-mass energy, $\sqrt{s}$, and integrates
\begin{equation}
\label{eq:int-datadriven}
a_{\mu}^{\mathrm{LOHVP}}=\frac{1}{4\pi^3}\int_{s_0}^{\infty} ds \sigma^0_{\mathrm{had,\gamma}}(s) K(s) \, .
\end{equation}
$\sigma^0$ is the bare cross-section with all hadron final states and final-state radiation included. Data from many experiments must be collated and averaged, taking correlations into account. At low values of $\sqrt{s}$ (below 2 GeV) data for exclusive final states are used, above this inclusive data, and at very high energies, QCD perturbation theory. Figure~\ref{fig:datadriven} shows a map of the low $\sqrt{s}$ data, which is emphasised by the kernel function. 75\% of the LOHVP result comes from $\sqrt{s}$ below 0.9 GeV~\cite{keshavarzi:2019abf}.  

The cross-section measurements can be made in two different ways, either by a direct `energy scan' at a fixed (but variable) colliding beam energy  (e.g. SND, CMD-3, KEDR) or by `radiative return' in which the beam energy does not change, but (tagged) initial-state photon radiation changes the collision energy (e.g. BaBar, BES III, KLOE). The tension between experimental results has to be taken into account and for WP20~\cite{Aoyama:2020ynm} a key issue was the tension between KLOE and BaBar for $\sigma(e^+e^-\to \pi^+\pi^-)$  below 1 GeV, a major contribution to the LOHVP. The two data-driven analyses that were combined for WP20 were from DHMZ19~\cite{davier:2019can} and KNT19~\cite{keshavarzi:2019abf}, as shown in Figure~\ref{fig:snowmass}. The latter dataset is publically available and has been widely used. 

\begin{figure}[thb]
  \centering
  \includegraphics[width=0.9\hsize]{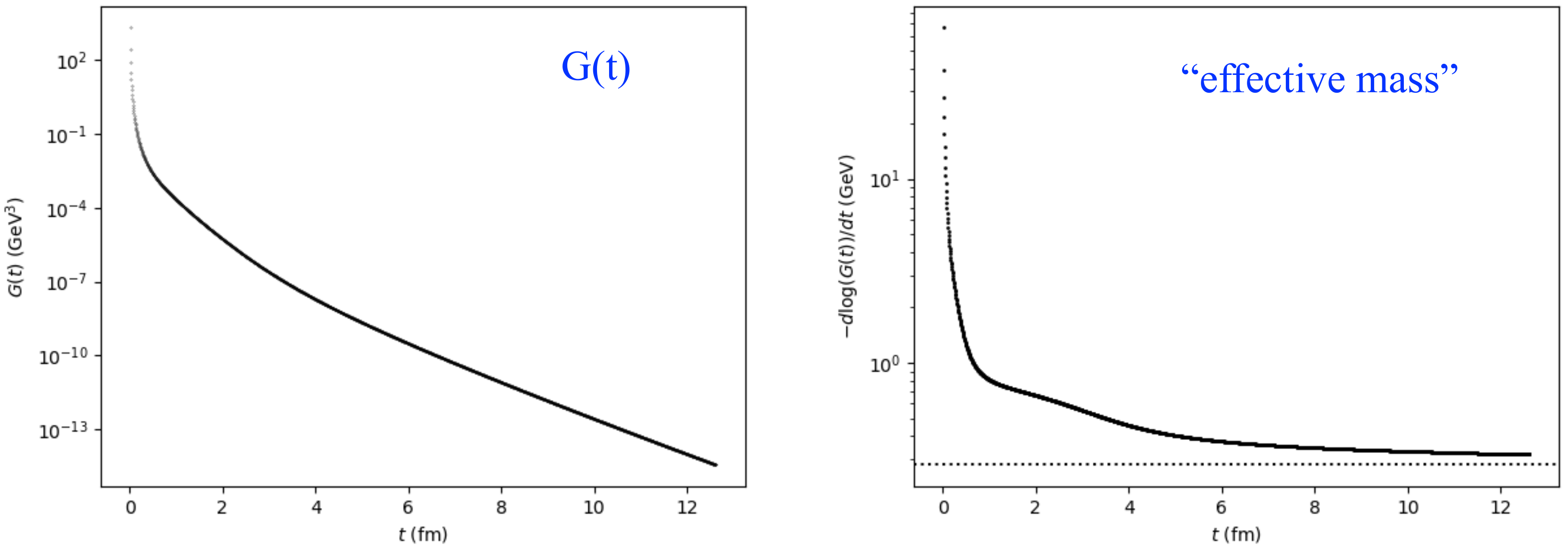}
  \caption{On the left, the correlator $G(t)$ determined from the $R(s)$ data in the KNT19 dataset~\cite{keshavarzi:2019abf} using Eq.~\eqref{eq:laplace}. This $G(t)$ unavoidably includes all quark flavours, along with disconnected contributions, QED etc. Note that the time-axis extends far beyond the $\sim 8$ fm that would be normal for a lattice QCD calculation. The plot on the right shows the effective mass determined from this correlator with the $\rho$ meson visible as a shoulder. Figures from P. Lepage. }
  \label{fig:KNT19G}% Give a unique label
\end{figure}

More recently, 2023 results from CMD-3~\cite{CMD-3:2023alj,CMD-3:2023rfe} for $\sigma(e^+e^-\to \pi^+\pi^-)$ have caused consternation by being higher than previous values and increasing the tension in this channel to the point where it is hard to see how to take an average (right-hand plot of Figure~\ref{fig:datadriven}). Further analysis will be needed to resolve this situation and work is underway; new results with improved statistics are expected from BaBar and KLOE, among others, in the near future. It has been suggested~\cite{Davier:2023fpl} that further attention should be paid to $\tau$ decay to hadrons via a virtual $W$. It is possible to select final-states from this decay that correspond to the vector piece of the $W$ $V-A$ current and this differs only from the QED current by an isospin rotation. The isospin corrections are complicated, however, and this mode was not used in WP20. There is an opportunity here for lattice QCD to make input; see the talk at this conference by M. Bruno. 

It is straightforward to compare lattice QCD and data-driven results in detail because the data-driven results can be converted from energy- to time-space using a Laplace transform~\cite{Bernecker:2011gh}.
\begin{equation}\label{eq:laplace}
G(t)= \frac{1}{12\pi^2}\int_0^{\infty} dE\, E^2 R(E^2) e^{-E|t|}
\end{equation}
where $E=\sqrt{s}$ and $R$ is the ratio of $\sigma(e^+e^-\to\mathrm{hadrons})$ to $\sigma(e^+e^-\to\mu^+\mu^-)=4\pi\alpha^2/(3s)$. Figure~\ref{fig:KNT19G} shows this transformation of the KNT19 dataset. Given a $G(t)$ from $R(s)$ we can apply the time-window functions in exactly the same way as was done for the lattice QCD results. Note that the partial LOHVP results are physical, so a disagreement between the lattice and data-driven values in any window is just as worrying as for the full LOHVP. 

\begin{figure}[t]
  \centering
  \includegraphics[width=0.9\hsize]{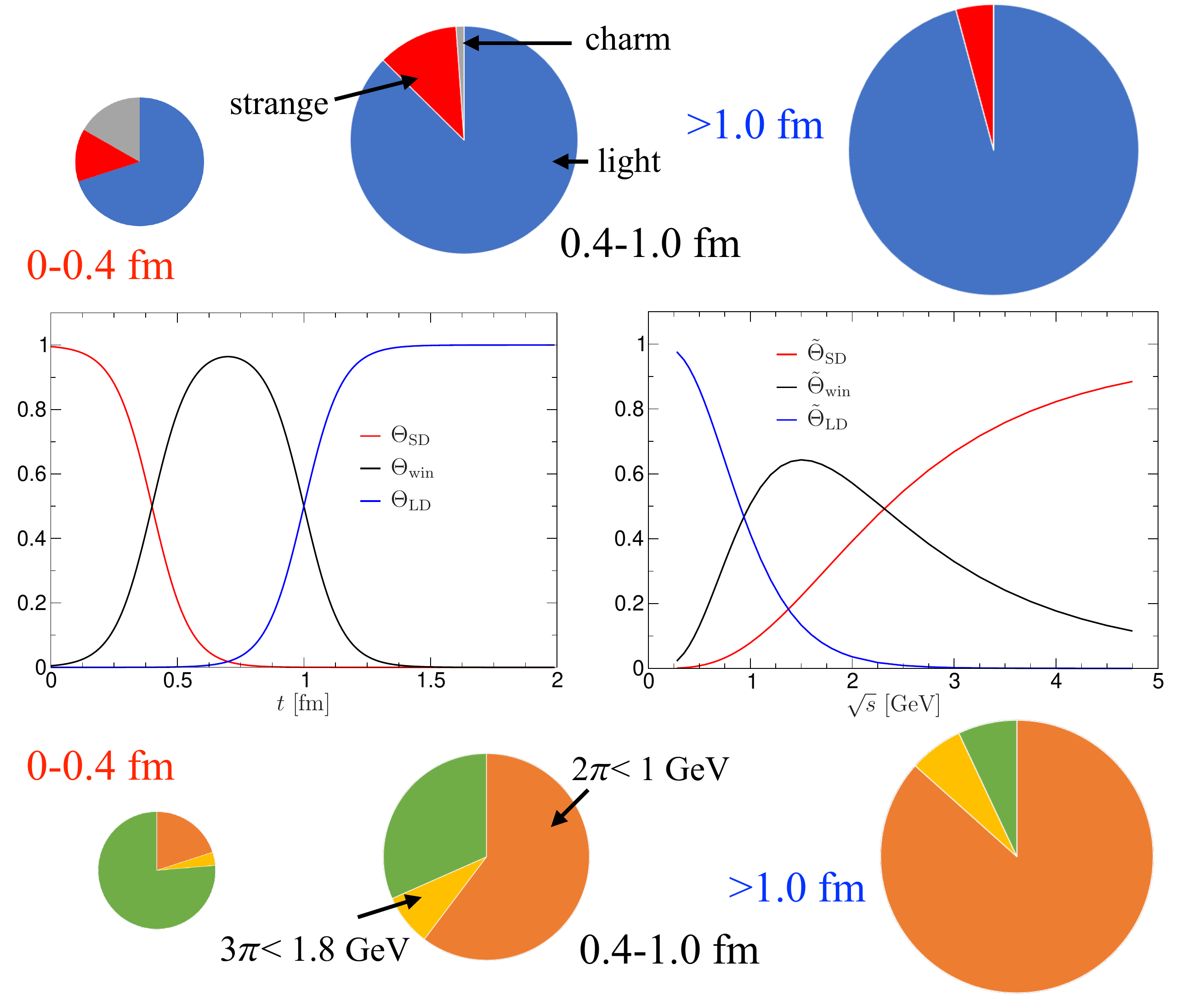}
  \caption{Perspectives on the standard time-windows. The central two plots show how the short-distance, intermediate distance (win) and long-distance time-windows map into modifications of the integrand viewed in $\sqrt{s}$ space. Figure from~\cite{Colangelo:2022vok}. The upper pie charts show the flavour contributions in the different windows from lattice results. The negative disconnected contribution is included under `light', but is only significant in the long-distance window. Data from~\cite{Djukanovic:2024cmq,FermilabLattice:2024yho}. The lower pie charts show how the makeup of the contributions from 3 different $e^+e^-$ channels to the integral varies between the time windows. The channels are $2\pi$ below 1 GeV (orange), $3\pi$ below 1.8 GeV (yellow) and the remainder (in green). We see that $2\pi$ < 1 GeV becomes increasingly dominant for the windows at larger time. Data from~\cite{Colangelo:2022vok}. The area of the pie charts is approximately in the ratio of the size of that time-window's contribution to the total (1:3.3:5.7). }
  \label{fig:Rswindows}% Give a unique label
\end{figure}

Figure~\ref{fig:Rswindows} shows the mapping of the short-distance, intermediate-distance and long-distance time windows into their equivalents in $\sqrt{s}$ space. As might be expected, the short-distance time window has a larger contribution from higher $\sqrt{s}$ values and the long-distance time window is completely dominated by $\sqrt{s} <1\,\mathrm{GeV}$. The lower pie charts show the varying makeup of the different time-windows in terms of 3 different hadronic channels covering different regions of $\sqrt{s}$~\cite{Colangelo:2022vok}. 

We can now compare results for different time-windows between lattice QCD and data-driven approaches. This is shown in Figure~\ref{fig:ID-comp}. The left-hand plot has results for the 0.4--1.0 fm windows from light-quark connected (lqc) correlators only and the right-hand plot shows 0--0.4 fm and 0.4--1.0 fm windows for the complete LOHVP calculation. The data-driven results can be compared directly to the complete LOHVP lattice results in the right-hand plot and the values shown come from experimental $e^+e^-$ data used in WP20. The lattice QCD results for 0.4--1.0 fm are clearly 3--4\% larger, with tension at the $4\sigma$ level. No tension is evident for the 0--0.4 fm window. This points to an issue in the $e^+e^-$ data at relatively low energy, following the lower pie charts of Figure~\ref{fig:Rswindows}. That this must be seriously considered is shown by the following fact. If the WP20 $e^+e^-$ dataset (KNT19) is modified so that CMD-3 data is substituted for the low-energy $e^+e^-\to \pi^+\pi^-$ cross-section then larger values for the windowed LOHVP are obtained that agree with the lattice QCD results. The plot on the left of Figure~\ref{fig:ID-comp} shows the same feature for the lqc case. Here the data-driven comparison is not as straightforward since other flavours and disconnected contributions must be `removed' from the $e^+e^-$ data~\cite{Benton:2023dci}. When this is done, a $6\sigma$ disagreement is seen with the BMW/DMZ24 lattice results~\cite{Boccaletti:2024guq}, with the same picture echoed by the earlier lattice QCD calculations.  When CMD-3 data is substituted, agreement is seen~\cite{Benton:2023fcv}. The long-distance window ($t >$ 1.0 fm) will be discussed in the next Section. 

\section{Update on the full LOHVP lattice results}
\label{sec:full-LOHVP}

\begin{figure}[t]
  \centering
  \includegraphics[width=0.9\hsize]{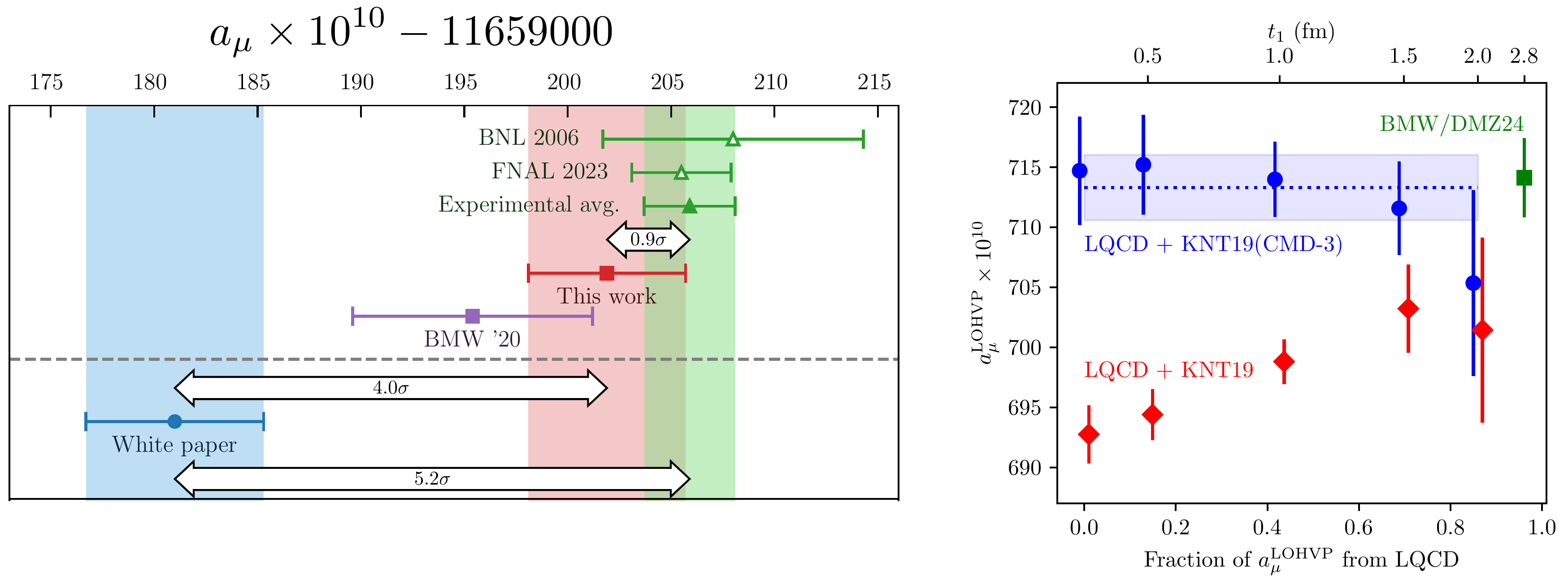}
  \caption{The left-hand plot is taken from~\cite{Boccaletti:2024guq} and shows the difference between SM theory and experiment for $a_{\mu}$ using the LOHVP from the new BMW/DMZ24 value (in red)  along with those from WP20~\cite{Aoyama:2020ynm} (blue) and BMW20~\cite{Borsanyi:2020mff} (purple). The experimental average (green triangle) is from~\cite{Muong-2:2023cdq}. BMW/DMZ24 combined lattice QCD results from a 0--2.8 fm time-window with a data-driven value for larger times. The plot on the right from~\cite{Davies:2024pvv} tests this hybrid approach as a function of $t_1$, the time point where the two are joined. A result independent of $t_1$, and in agreement with BMW/DMZ24,  is found when CMD-3 $\pi^+\pi^-$ data is substituted in the KNT19 dataset. This value also agrees with the pure data-driven result using this KNT19(CMD3) dataset (left-most blue point).}
  \label{fig:hybrid}% Give a unique label
\end{figure}

Just before the Lattice2024 conference  the BMW/DMZ collaboration released a new result for the full LOHVP~\cite{Boccaletti:2024guq} of $714.1(3.3)\times 10^{-10}$, 3\% higher than the WP20 value of Eq.~\eqref{eq:contribs} and with a 0.5\% uncertainty. The picture that this gives when compared to the experimental $a_{\mu}$ result (taking all other contributions from WP20~\cite{Aoyama:2020ynm}) is shown on the left in Figure~\ref{fig:hybrid}. It supports a `no new physics in $a_{\mu}$' scenario. BMW/DMZ24 improves the uncertainty over the BMW20 result~\cite{Borsanyi:2020mff} by adding additional lattice data from a finer ensemble but also by using a 0--2.8 fm window for the lattice data and adding a value from data-driven results for 2.8--$\infty$ fm. This latter move reduces statistical errors and finite-volume corrections (and hence uncertainties) for the lattice data. The downside is that it then includes a small, but not negligible, contribution ($27.6\times 10^{-10} =$ 3.9\%) based on $e^+e^-$ data that have been called into question (see Figures~\ref{fig:ID-comp} and~\ref{fig:datadriven}). How much should we worry about this? This was a very lively topic of conversation at the Lattice2024 conference. 

The question calls for a more systematic study of the hybrid approach (combining lattice QCD and data-driven results) for determining the full LOHVP. This is done in~\cite{Davies:2024pvv} by adding lattice QCD results for a time-window from 0--$t_1$ with data-driven results from $t_1$--$\infty$, as a function of $t_1$. As well as giving a total LOHVP, this method tests the consistency between lattice QCD and data-driven results; if they are consistent you should not be able to see the `join' at $t_1$ and should get the same result independent of $t_1$. This exercise is shown in the right-hand plot of Figure~\ref{fig:hybrid} using Fermilab/HPQCD/MILC lattice data from 2019~\cite{FermilabLattice:2019ugu}. We see that a flat total is only obtained when CMD-3 $\pi^+\pi^-$ data is substituted in KNT19, and this is backed up by a fully correlated fit to a constant (dashed blue line and band on the plot). The resulting total LOHVP in this case agrees with that from BMW/DMZ24. 

Much better results could be obtained in this approach using newer lattice data and it has the potential to provide a more accurate total LOHVP than either lattice or data-driven results on their own (as Figure~\ref{fig:hybrid} indicates). The argument that data-driven results should now be completely discarded seems an unnecessarily extreme one. As we have seen, lattice QCD numbers provide strong support for the CMD-3 data. In the absence of an understanding of the difference with earlier data, however, a systematic uncertainty that takes into account this difference might be reasonable. A key point for the hybrid approach is that for $t_1$ values that are large, but still accessible on the lattice with good statistical accuracy, the difference in the total between using KNT19 and using KNT19(CMD-3) is small compared to the total uncertainty (at least for current lattice results). For $t_1=2.5$ fm, the difference is only $1.7\times 10^{-10}$. 

Other collaborations discussed their pure lattice QCD results for the light-quark connected contribution to the long-distance window and hence, combining with the other windows, the full lqc piece of the LOHVP. The RBC/UKQCD collaboration unblinded their results for the lattice conference and they were presented by C. Lehner~\cite{RBC:2024fic}; the Mainz/CLS~\cite{Djukanovic:2024cmq} and Fermilab/HPQD/MILC~\cite{Bazavov:2024eou} results were unblinded soon after.  ETMC will also have a result shortly; in the meantime they have updated their $s$ and $c$ results~\cite{ExtendedTwistedMassCollaborationETMC:2024xdf}. 

\begin{figure}[t]
  \centering
  \includegraphics[width=1.0\hsize]{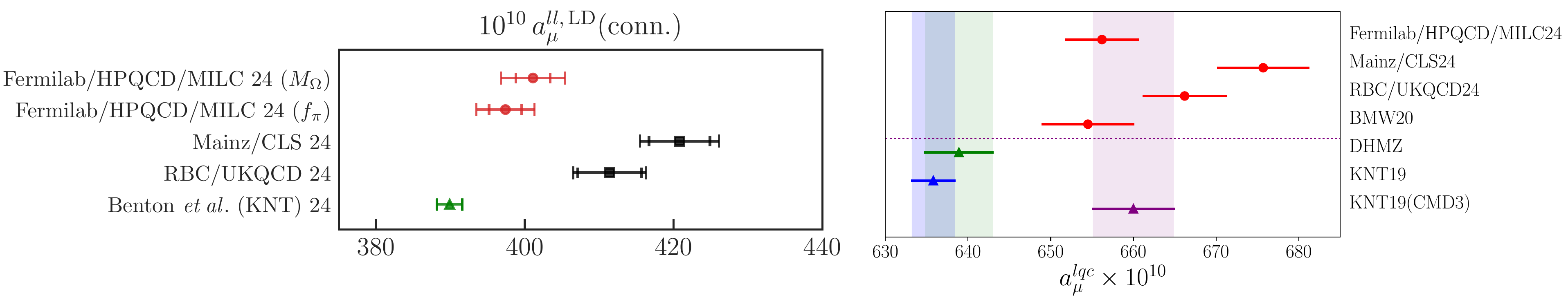}
  \caption{The left-hand plot is taken from Fermilab/HPQCD/MILC24~\cite{Bazavov:2024eou} and shows a comparison of lattice results for the light-quark connected contribution in the long-distance window, for $t >$ 1.0 fm. The top two points are from Fermilab/HPQCD/MILC using different quantities to fix the lattice spacing. The other lattice results are from Mainz/CLS~\cite{Djukanovic:2024cmq} and RBC/UKQCD~\cite{RBC:2024fic}, using different quark formalisms. The green filled triangle is based on the KNT19 $e^+e^-$ dataset~\cite{Benton:2024kwp}. When CMD-3 $\pi^+\pi^-$ data is substituted in the KNT19 dataset, the value increases to 406.5 in an exploratory analysis~\cite{Benton:2024kwp}, closer to the lattice results. \\On the right is a comparison of the full light-quark connected contribution when each collaboration adds the long-distance (Fermilab/HPQCD/MILC use their $M_{\Omega}$ value) to the short-distance and intermediate-distance windows. Also shown is the value from BMW20~\cite{Borsanyi:2020mff} (connected light contribution +10/9 of the finite-volume correction, splitting this between connected and disconnected in the ratio of the $\pi\pi$ contributions~\cite{Chakraborty:2015ugp}). Below the dotted line are three data-driven results from~\cite{Benton:2024kwp}: the green point and band use DHMZ, the blue use KNT19 and the purple point KNT19(CMD3). }
  \label{fig:LDlqc}% Give a unique label
\end{figure}

The long-distance window is the hardest, with the largest lattice statistical and systematic uncertainties, as discussed in Section~\ref{sec:latwindows}. The new lattice results have uncertainties at the level of 1.0\% which represents a big improvement on what has previously been possible. The results are shown on the left of Figure~\ref{fig:LDlqc}. There is some spread/tension in the values, reflecting the difficulty of calculating in this window. More work may be needed to understand this. All of the values are above the data-driven result determined using KNT19 and correcting for other flavour contributions. 

The right-hand plot of Figure~\ref{fig:LDlqc} shows the light-quark connected contribution integrated over the full time range for the three lattice results from the left-hand plot as well as the BMW20 result. The same tensions visible in the left-hand plot are still there. The coloured bands correspond to data-driven values and it is clear that the best agreement with lattice QCD results is for the dataset in which CMD3 data has been substituted in the KNT19 dataset, agreeing with the narrative we have seen from other results. 

The Mainz/CLS collaboration combine their long-distance window with their previous results, including for other flavours and QED and isospin-breaking effects, to quote a new value for the total LOHVP of $724.9(7.0)\times 10^{-10}$~\cite{Djukanovic:2024cmq}, slightly higher than the BMW/DMZ24 value of $714.1(3.3)\times 10^{-10}$~\cite{Boccaletti:2024guq} but also giving an $a_{\mu}$ value consistent with the experimental average,  

As remarked earlier, these lattice LOHVP numbers are significantly larger than the data-driven values included in WP20. At the time that the similarly large LOHVP result from BMW20 appeared~\cite{Borsanyi:2020mff}, there were worries that this would imply, through their common dependence on the hadronic vacuum polarisation function, a larger hadronic contribution to the running of $\alpha$ that would conflict with values of $\Delta\alpha_{\mathrm{had}}(M_Z)$ from electroweak precision fits. However, because of the different energy-sensitivity in the two calculations this does not happen~\cite{Borsanyi:2020mff,Ce:2022eix} and shifts are within the level of uncertainty on $\Delta\alpha_{\mathrm{had}}(M_Z)$. This is true also if CMD3 data is substituted in the KNT19 $e^+e^-$ dataset~\cite{DiLuzio:2024sps}.  MuonE@CERN~\cite{MUonE:2016hru} is an experiment that aims to map out the running of $\alpha$ using $\mu-e$ scattering and determine its hadronic component to better than 0.5\%, then also weighing in on the LOHVP contribution to $a_{\mu}$. Test runs have begun but the full apparatus is due to be installed after the long--shutdown from 2026-29. 

Given that the HOHVP contributions (see Section~\ref{sec:QCD}) use the same QCD input as the LOHVP but sum to a total ($\mathcal{O}(\alpha^3)+\mathcal{O}(\alpha^4)$) that is 1.2\% of the size, the issues with the LOHVP will have little impact there. We now turn to the other important QCD contribution that must be pinned down: the hadronic light-by-light. This was reviewed at Lattice2023 by A. Gerardin. 

\section{The HLbL contribution} 
\label{sec:hlbl}
As for the LOHVP, there are two methods to determine the HLbL contribution. However in this case the data-driven, dispersive approach is less direct and the calculation must be broken down into multiple pieces. The contribution can instead be calculated directly in lattice QCD. It is an enormously complicated calculation but, following the pioneering work of RBC/UKQCD and Mainz/CLS, it is now tractable. The two approaches give a consistent picture and WP20~\cite{Aoyama:2020ynm} combined the results~\cite{Melnikov:2003xd,Masjuan:2017tvw,Colangelo:2017fiz,Hoferichter:2018kwz,Gerardin:2019vio,Bijnens:2019ghy,Colangelo:2019uex,Pauk:2014rta,Danilkin:2016hnh,Jegerlehner:2017gek,Knecht:2018sci,Eichmann:2019bqf,Roig:2019reh,Blum:2019ugy} to quote a value of $9.0(1.7)\times 10^{-10}$. 

\begin{figure}[t]
  \centering
  \includegraphics[width=1.0\hsize]{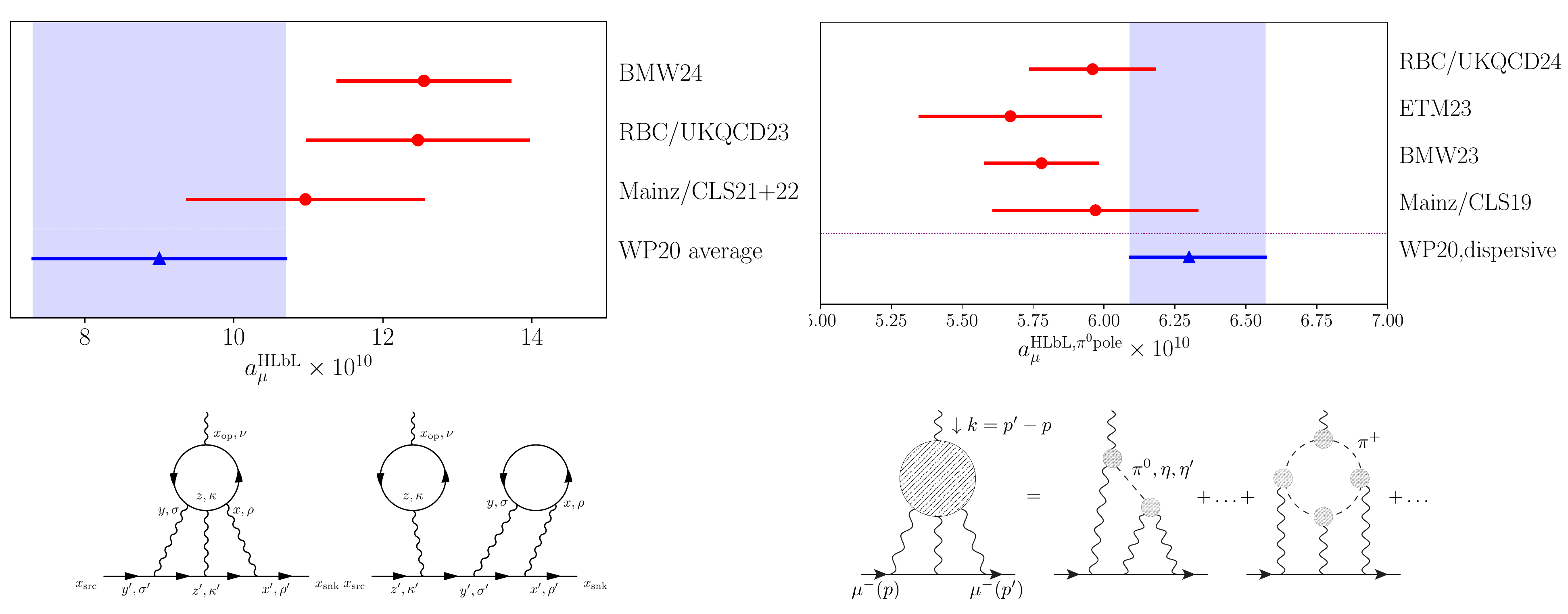}
  \caption{The left-hand plot compares direct lattice QCD calculations of the HLbL contribution from 4-point correlators sketched in the diagram below. The lattice results in red use 3 different quark formalisms: BMW24~\cite{Fodor:2024jyn}, staggered quarks; RBC/UKQCD23~\cite{Blum:2023vlm}, domain-wall; Mainz/CLS~\cite{Chao:2021tvp,Chao:2022xzg}, Wilson, and agree well. The RBC/UKQCD superseded value from 2019~\cite{Blum:2019ugy} is not shown but also agrees. The value used in WP20 is shown by the blue point and band. \\
  The right-hand plot compares lattice results for the $\pi^0$-pole piece of the HLbL contribution. This is the largest single piece, as motivated by the diagram below. The lattice results in red are from 4 different quark formalisms~\cite{Gerardin:2019vio,Gerardin:2023naa,ExtendedTwistedMass:2023hin,Lin:2024khg} and are compared to the WP20 value for this piece, within the dispersive framework, shown in blue. }
  \label{fig:HLbL}% Give a unique label
\end{figure}

 The direct lattice calculation involves a 4-point correlation function that must be evaluated for light, $s$ and $c$ quarks, along with quark-line disconnected pieces.  Figure~\ref{fig:HLbL} shows the current status of the lattice results on the left with a diagram below showing the connected and leading disconnected contributions. At this conference a new result was presented by C. Zimmermann for the BMW collaboration using staggered quarks~\cite{Fodor:2024jyn}. This agrees well with earlier results from RBC/UKQCD~\cite{Blum:2023vlm}  and Mainz/CLS~\cite{Chao:2021tvp,Chao:2022xzg} using different formalisms, as shown in Figure~\ref{fig:HLbL}. Preliminary results for the ETM collaboration using twisted-mass quarks were also shown at the conference by N. Kalntis~\cite{Kalntis:2024dyd}.

Figure~\ref{fig:HLbL} shows that the post-WP20 lattice results have central values slightly higher than the WP20 value and somewhat smaller uncertainty at around 10\%. This probably indicates that the HLbL contribution is slightly larger than the WP20 value, but only by a very small shift, $\sim 3\times\,10^{-10}$. This is around the size of the uncertainty of the LOHVP contribution discussed in Section~\ref{sec:full-LOHVP}. Hence any change of this size will have minimal impact on the total SM value for $a_{\mu}$. This is very reassuring to know, and lattice QCD is the only way to provide this level of certainty about the numbers in such a direct way. 

The dispersive approach to the HLbL builds that contribution from low-energy single meson exchange plus loops with charged pions and kaons, followed by perturbation theory when all momenta are large (see the diagram on the right of Figure~\ref{fig:HLbL}). The meson exchanges that dominate are those of the light pseudoscalars, $\pi^0$, $\eta$ and $\eta^{\prime}$, but there are also contributions from heavier scalar, axial-vector and tensor mesons. The pseudoscalar-meson-pole contributions, which are numerically the most important, can be written in terms of transition form factors, $F_{P\gamma^*\gamma*}(Q_1^2,Q_2^2)$. These are experimentally accessible for different $Q^2$ regions from two-photon collisions (at $e^+e^-$ colliders), Dalitz decay ($P\to e^+e^-\gamma$), and $e^+e^-\to \gamma P$~\cite{Aoyama:2020ynm}. A test of the values at $Q_1^2=Q_2^2=0$. comes from comparing $\Gamma(P\to\gamma\gamma)$ to experiment. 

The transition form factors can also be calculated in lattice QCD. The $\pi^0$ pole contribution has now been determined by four collaborations using different quark actions; a new result from RBC/UKQCD was presented at the conference by Tian Lin~\cite{Lin:2024khg} and it agrees well with the earlier values as shown on the right in Figure~\ref{fig:HLbL}. The lattice results agree reasonably well with the WP20 value for this piece, shown by the blue band, although they are all on the low side of it. The BMW23 calculation~\cite{Gerardin:2023naa} includes also the $\eta$ and $\eta^{\prime}$ contributions, obtaining a total for the pseudoscalar poles of $8.51(52)\times 10^{-10}$ to be compared to $9.38(40)\times 10^{-10}$ from WP20. 
The lattice QCD results therefore provide significant validation of the pseudoscalar pole contributions in this dispersive framework. Although these are the largest, they are not the only contribution; there are more pieces, of size $\sim 1.5\times 10^{-10}$, and they have varying signs. There is nothing, however, to suggest any major disagreement between the direct lattice results and the dispersive approach for the HLbL contribution and this gives significant added confidence in theorists' ability to determine this.

\section{Conclusions}
We have seen huge progress on lattice QCD results for the QCD contributions to $a_{\mu}$ since the 2020 Theory White Paper~\cite{Aoyama:2020ynm}. The lattice is now pinning down the largest such contribution, the LOHVP. Accurate results have been demonstrated for time-windowed pieces, with impressive agreement for the 0.4--1.0 fm window from many independent calculations with different quark actions (Figure~\ref{fig:ID-comp}). This is a major achievement for the lattice. There are also now several lattice results for the full light-quark connected contribution (Figure~\ref{fig:LDlqc}) and two new values for the complete LOHVP (Section~\ref{sec:full-LOHVP}). There has been continuing important lattice progress on the HLbL contribution to $a_{\mu}$ (Section~\ref{sec:hlbl}), without raising any surprises. 

The lattice values for the LOHVP, in contrast, call into question the data-driven numbers included in WP20 using the $e^+e^-$ experimental data available at that time. There is significant disagreement between the lattice and WP20 data-driven results, particularly for the 0.4--1.0 fm window. This is shown very graphically when lattice and data-driven results are combined (Figure~\ref{fig:hybrid}, right). Instead the lattice QCD results show a clear preference for the CMD3 $e^+e^-\to\pi^+\pi^-$ data over earlier experimental results for this channel. It is critical that the differences within the experimental $e^+e^-$ data should be understood; it is to be hoped that the new experimental analyses underway will shed light on this. 

Lattice QCD results for the partial and full LOHVP all now point towards an LOHVP contribution to $a_{\mu}$ which is $\sim$3--4\% larger than that given in WP20. This implies an SM $a_{\mu}$ value which is much closer to that seen in experiment, with the conclusion that there is much less new physics visible in $a_{\mu}$ than we had hoped, and perhaps none. There is still a lot to be done before we can arrive firmly at that conclusion. We need multiple lattice QCD values for the full LOHVP aiming at 0.5\% precision from different groups using blinded analyses. This is well underway. Some improvements are still needed to reach this accuracy from a purely lattice calculation and a hybrid approach combining lattice QCD with data-driven values for the large-time tail should be considered. 

Looking ahead, the final result for $a_{\mu}$ from the Muon g-2 experiment is expected in Spring 2025, preceded by a new Theory White Paper. A further measurement of $a_{\mu}$ is planned at J-PARC@KEK~\cite{Abe:2019thb} using a very different technique with a compact magnetic ring and low-momentum $\mu^+$. Data-taking should start in 2028 with 2 years of running needed to reach a result with $\sim 2\times$ the uncertainty of Muon g-2. 

\vspace{2mm}
{\bf Acknowledgements} \\
I am grateful to the organisers for giving me the opportunity to review this topic. I have had a lot of useful input from A. Keshavarzi and P. Lepage and also benefitted from discussions with A. El-Khadra, F. Erben, M. Garofolo, D. Giusti, M. Golterman, Luchang Jin, N. Kalntis, A. Kotov, A. Kronfeld, S. Kuberski, S. Lahert, C. Lehner, Tian Lin, C. McNeile, L. Parato, T. Teubner, U. Wenger and C. Zimmermann. This has been a most interesting and enjoyable conference and I thank all the participants for that. 

\bibliography{muong2}{}
\bibliographystyle{JHEP}

\end{document}